\documentclass[12pt,usenatbib,useAMS]{iopart}
\usepackage[pdftex]{hyperref}
\usepackage[pdftex]{graphicx}

\begin{document}
\title{A graphical tool for demonstrating the techniques of radio interferometry}
\author{A. Avison$^1$ and S. J. George$^2$}
\address{$^1$ UK ALMA Regional Centre Node, Jodrell Bank Centre for Astrophysics, Alan Turing Building, University of Manchester, Manchester, M13 9PL, UK.}
\address{$^2$ Astrophysics Group, Cavendish Laboratory, University of Cambridge, Cambridge, CB3 0HE, UK.}
\ead{adam.avison@manchester.ac.uk}

\begin{abstract}
We present a graphical interface designed to demonstrate the techniques of radio interferometry used by telescopes like ALMA, e-Merlin, the JVLA and SKA, in a manner accessible to the general public. Interferometry is an observational technique used by astronomers to combine the signal from a few to tens to hundreds of individual small antennas to achieve high resolution images at radio and millimetre wavelengths. This graphical interface demonstrates how the number of antenna, their position relative to one another and the rotation of the Earth allow astronomers to create highly detailed images at long wavelengths. \\

\noindent -\textit{Accepted for publication in the European Journal of Physics}\footnote{This is an author-created, un-copyedited version of an article accepted for publication in European Journal of Physics. IOP Publishing Ltd is not responsible for any errors or omissions in this version of the manuscript or any version derived from it. The Version of Record is available online at \href{dx.doi.org/10.1088/0143-0807/34/1/7}{doi:10.1088/0143-0807/34/1/7}.}. 
\end{abstract}



\section{Introduction}
The technique of interferometry is widely used in the field of radio astronomy, with some of the worlds largest scientific endeavours based upon it (e.g. the Atacama Large Millimeter/Sub-mm Array\cite{ALMAObsWebsite, ALMA2004} and the up coming Square Kilometre Array\cite{SKAWebsite, SKADesign}). Over the coming years telescopes such as these will revolutionise astronomy and hopefully inspire a new generation of students to take an interest in science. With this in mind the need to explain the techniques of interferometry beyond the most basic level becomes apparent, however a complete treatment of these techniques requires an understanding of several areas of mathematics beyond that taught prior to undergraduate level, namely Fourier transforms and complex numbers. For completeness we have included in this paper such mathematics and its application to interferometry,  we aim to have presented this in such a way that the paper is accessible to readers part way through an undergraduate Physics degree (e.g. 2$^{nd}$ year and higher) and beyond. 

Whilst these areas of mathematics are vital for an interferometer they are required only in the `intermediate' stage of interferometric data processing, the stage in which the signals received at each individual antenna are combined to produce the final image.  

To this end we have developed a graphical tool to demonstrate some of the real life effects on an interferometric image of antenna position and number and Earth Rotation synthesis, without outwardly discussing the mathematics required at the intermediate stage to generate the final image. The tool itself is written in the Python\footnote{see \texttt{python.org}} scripting language and can be run on Windows, Mac and Linux. The tool is called the ``Pynterferometer", being that it is a Python based interferometer. 

This paper is organised as follows; first we address why astronomers use interferometry, and then present the fundamental equations used in the derivation of a final image from an observation and the effect of increasing the number of antennas. The layout of an array and the use of Earth's rotation in creating a high quality observation are then discussed. The approach used by our graphical tool is then discussed and an outline of the user interface and example demonstrations are outlined. Finally the reception of the tool by the general public is discussed.



\section{Why use interferometry?} For a conventional telescope the maximum resolution is approximately described by:
\begin{equation}
\theta = 1.22 \frac{\lambda}{D} 
\label{singlesres:eqn}
\end{equation}

\noindent where $\theta$ is the angular resolution in radians\footnote{$\theta$ is typically referred to in arcseconds and can be converted from radians by multiplication by $\frac{180\times3600}{\pi}$}, $\lambda$ is the wavelength and $D$ the diameter of the mirror/parabolic dish. For optical wavelengths ($\lambda\sim$10$^{-7}$m) the mirrors required to produce angular resolutions of milliarcseconds need only be a few tens of meters in diameter. However, in the mm/radio regime, with wavelengths $\lambda\sim$10$^{-4}$ - 10$^{-3}$m, antennas would need parabolic dishes of $\sim$ 10$^4$m to 10$^7$m.  The largest steerable single dish radio telescope have diameters of up to 100m, beyond this major feats of engineering (not to mention finance) are required.

So in order to achieve resolutions comparable with optical telescopes the field of mm/radio astronomy instead uses the technique of interferometry, where the signals received by two or more antennas are combined together to approximate having a single large dish. The maximum resolution of an interferometer is described by:

\begin{equation}
\theta \approx \frac{\lambda}{B_{max}}
\label{interRES:eqn}
\end{equation}

\noindent where the symbols have the same meaning as Equation \ref{singlesres:eqn} but now $D$ has been replaced by $B_{max}$ the maximum separation, or baseline, between two antennas of the interferometer. The larger the longest baseline the greater the achievable resolution.
 

\section{A basic interferometer}
\begin{figure}[!h]
\begin{center}
\includegraphics[scale=0.5]{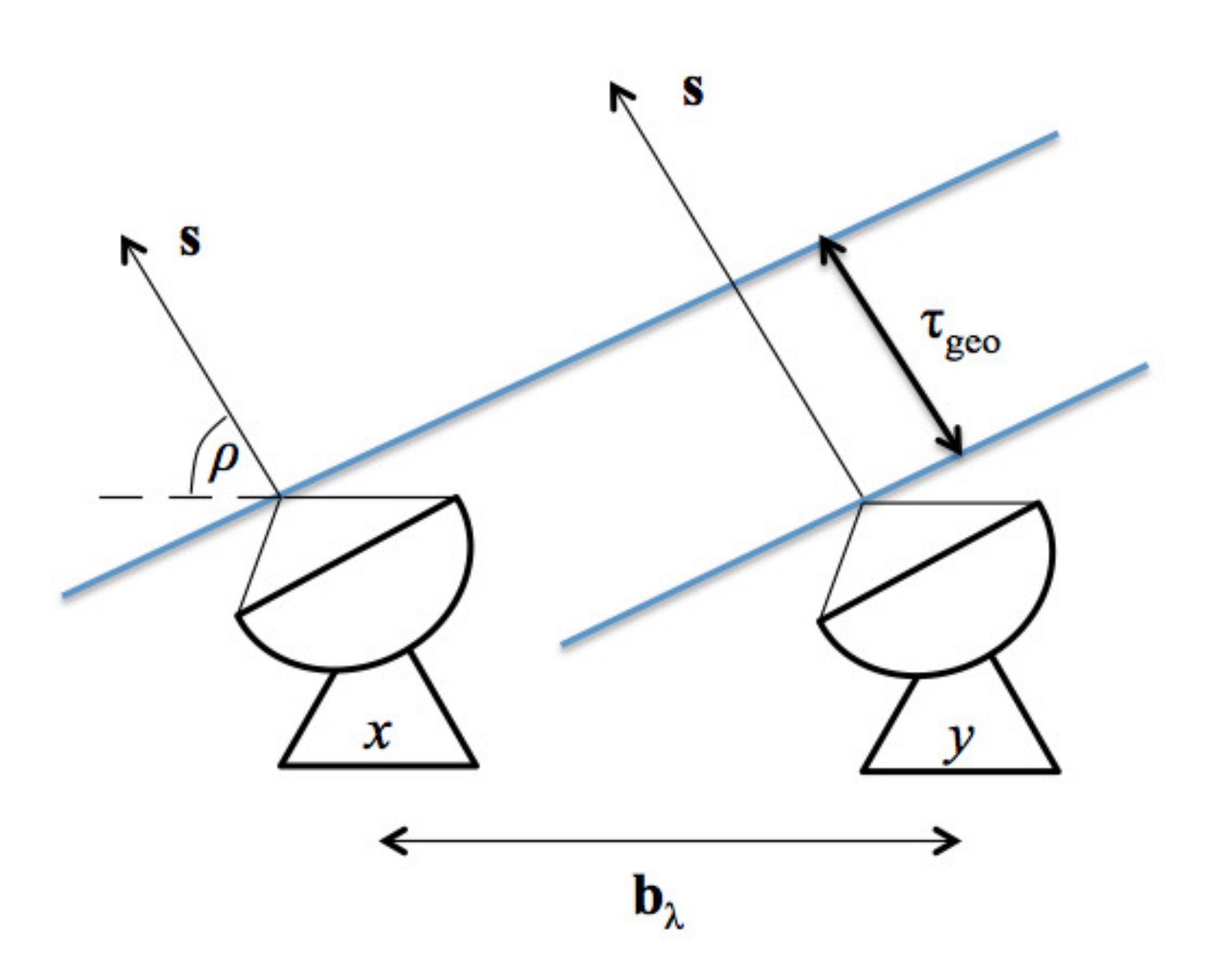}
\caption{Diagram of basic two element interferometer. Blue lines represent the incoming wavefront of photons at times $t$ and $t+\tau_{geo}$.}
\label{twoants:fig}
\end{center}
\end{figure}

The simplest case to consider when describing an interferometer is one consisting of only two antennas. In this section we follow the analysis of such a system by \cite{SynthesisImagingII} and \cite{IntroRadio}. Figure \ref{twoants:fig} shows the geometric setup of a two antenna interferometer and the rays of an incoming astronomical signal.


Consider photons arriving in a plane wavefront at our two antennas from an astronomical object, with sky brightness distribution $I$; the wavefront will be incident on antenna $x$ before antenna $y$. With the time delay between the wavefront arrival at $x$ and $y$ given by:

\begin{equation}
\tau_{geo} = \frac{\textbf{b}\cdot\textbf{s}}{c} = \frac{bs\cos\rho}{c}
\end{equation}

\noindent where \textbf{b} is the baseline length in metres, \textbf{s} a unit vector in the direction of the astronomical source, $\rho$ the angle between $b$ and $s$, and $c$ the speed of light. This delay is known as the geometric time delay as its existence is entirely due to the geometry of the system and the finite speed of light. The signals out of each antenna are of the form:
\begin{equation}
x(t)=v_{1}\cos2\pi\nu t
\end{equation}
\begin{equation}
y(t)=v_{2}\cos2\pi\nu (t+\tau_{geo})
\end{equation}

\noindent where $v_{1}, v_{2}$ are the signal voltage from each antenna, $\nu$ the observing frequency and $t$ time. The signals from each dish are fed into a computer known as a \textit{correlator} where the signals are cross correlated, (denoted by $\otimes$):


\begin{equation}
R_{x,y}(\tau_{geo})=x \otimes y= x(t) * y(-t)=X(\nu)Y^{*}(\nu)
\end{equation}

\noindent here $*$ denotes the convolution of signals $x(t)$ and $y(t)$ (with a time reversal for signal $y(t)$). This convolution is identical to the multiplication of the Fourier transform of $x(t)$ denoted $X(\nu)$ with the complex conjugate of the transform of $y(t)$ denoted $Y^*(\nu)$. This ultimately gives us a correlator output:

\begin{equation}
R_{x,y}(\tau_{geo})=v_1v_2\cos2\pi\nu\tau_{geo}\
\label{correlatorout:eqn}
\end{equation}

\noindent As we are observing astronomical signals the voltage outputs of antennas $x$ and $y$, ($v_1, v_2$), are related directly to the brightness of the astronomical object, $I(\textbf{s})$, over a solid angle d$\Omega$ ``seen" by our telescope and the area of the each dish we use to observe it, $A(\textbf{s})$. Leading Equation \ref{correlatorout:eqn} to become:

\begin{equation}
R_{x,y}(\tau_{geo})=\Delta\nu\int A(\textbf{s})I(\textbf{s})\cos2\pi \textbf{b}_{\lambda}\cdot\textbf{s} \quad d\Omega
\label{correlatorou2t:eqn}
\end{equation}

\noindent with \textbf{b}$_{\lambda}$ being the baseline length measured in wavelengths of the observations (i.e \textbf{b} in metres divided by $\lambda$). Radio/mm telescopes do not observe at a single unique frequency but instead cover a number of adjacent frequencies, known as a bandwidth, this is symbolised as $\Delta\nu$. 

As an observation made with an array telescope contains information from not just a single point but an area of the sky (the solid angle denoted by d$\Omega$), a reference point on the sky known as the \textit{phase tracking centre} is used within the observations. This is the point along vector \textbf{s$_0$}, with all other observed points at offset vectors \textbf{$\sigma$} such that \textbf{s}=\textbf{s$_0$}+\textbf{$\sigma$}. For the phase tracking centre an instrumental time delay is introduced to make $\tau_{geo}$ = 0 meaning all time delays in the signals are due to the positional offset \textbf{$\sigma$}.


Finally we define a function known as the \textit{complex visibility} as,

\begin{equation}
V=|V|e^{i\phi_{V}}=\int \mathcal{A}(\mathbf{\sigma})I(\mathbf{\sigma})e^{-i2\pi\mathbf{b}_{\lambda}\cdot\sigma}\quad d\Omega
\label{complexvisibility:eqn}
\end{equation}

\noindent where $\mathcal{A}(\sigma)$ indicates we have normalised, by a factor A$_0$, the antenna responses within the array and all other symbols have the meaning previously described. Substituting Equation \ref{complexvisibility:eqn} and the vector addition, \textbf{s}=\textbf{s$_0$}+\textbf{$\sigma$}, into Equation \ref{correlatorou2t:eqn} we are left with,

\begin{equation}
R_{x,y}=A_0|V|\Delta\nu\cos(2\pi\mathbf{b}_{\lambda}\cdot\mathbf{s}_0-\phi_V)
\label{correlatorou3t:eqn}
\end{equation}

\noindent representing the correlator output. From which we can recover the unknowns which are directly related to the source, $|V|$ and $\phi_V$, by Fourier transform of this value to obtain a component of the sky brightness distribution $I(\sigma)$.

\subsection{A convenient coordinate frame}

\begin{figure}[!h]
\begin{center}
\includegraphics[scale=0.7]{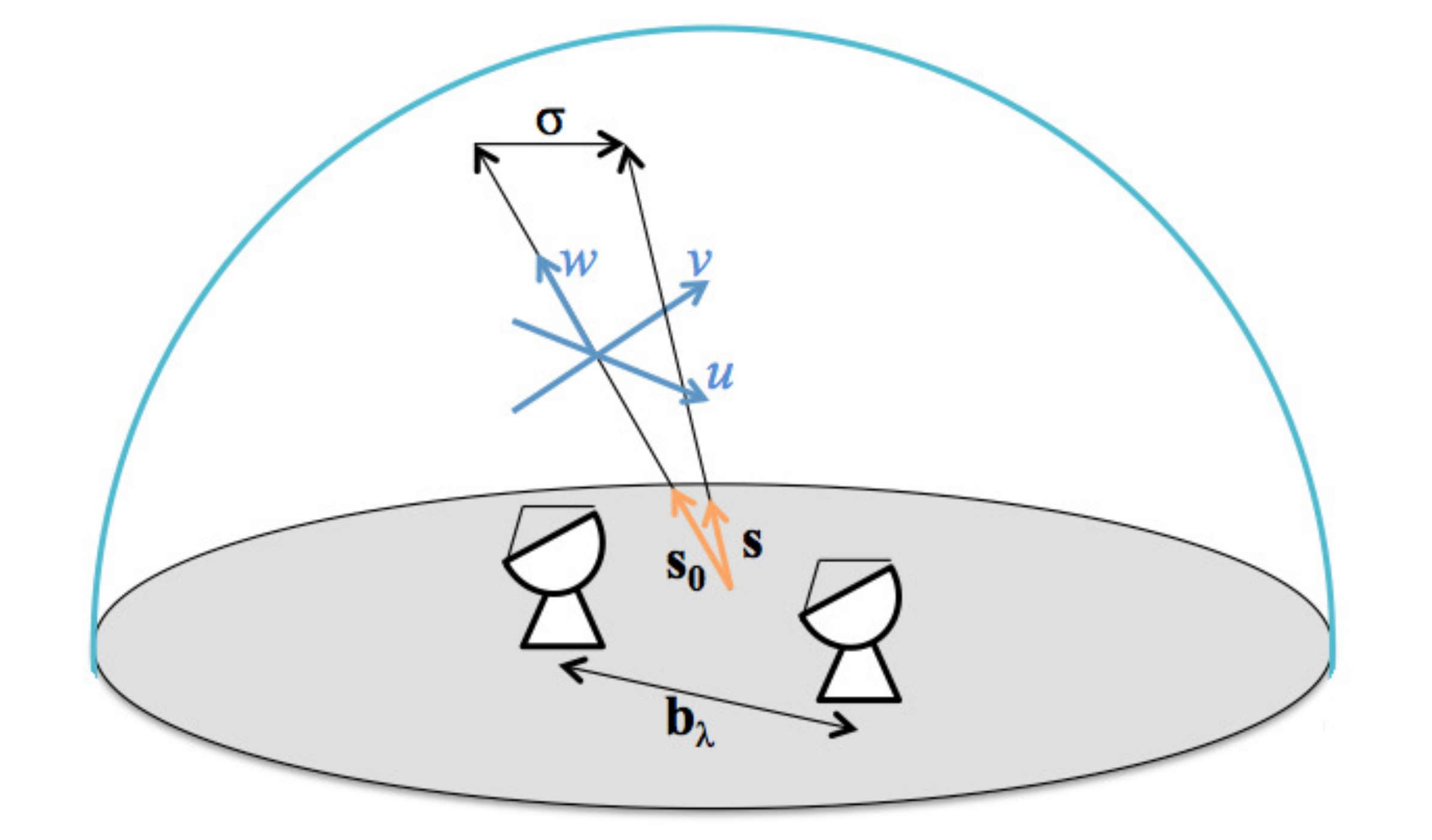}
\caption{Diagram of vectors and coordinate systems used in Equations from the text.}
\label{uvplane:fig}
\end{center}
\end{figure}

For interferometric observations there is a convenient coordinate frame within which to calculate the $complex$ $visibility$, which is key to the function of the Pynterferometer. We define the coordinate $w$ along vector \textbf{s}, with a plane perpendicular to $w$ defining the east-west coordinate, $u$, and the north-south coordinate, $v$. Figure \ref{uvplane:fig} represents this graphically.  As, in this frame, the offset vector $\sigma$ is parallel to the $uv$-plane it can be defined in $u, v$ coordinates, giving Equation \ref{complexvisibility:eqn} as,

\begin{equation}
V=\int \mathcal{A}(l,m)I(l,m)e^{-i2\pi(ul+vm)}\quad \frac{dl\,dm}{\sqrt{1-l^2-m^2}}
\label{complexuvvisibility:eqn}
\end{equation}

\noindent where $l$ and $m$ are directional cosines of the vector \textbf{s}. These can typically be replaced with the actual angular offsets ($i$ and $j$) using the small angle approximation (owing to the small angles subtended by $l$ and $m$). So $V(u, v)$ is then simply the Fourier transform of $I(i,j)$ the sky brightness distribution of the target observation.\\


\subsection{Number and positions of antennas}

From the above mathematical approach we can see that each pair of antennas for each time interval of observation gives a complex visibility measurement at a single $u, v$ point (plus its complex conjugate point). To recover the optimal amount of information about an astronomical object observation at as many $u, v$ points as possible are desirable.  For an interferometer of $N$ antennas there are $N(N-1)/2$ antenna pairs and therefore $N(N-1)$ $u, v$ points in a single measurement. With each antenna pair acquiring spatial information of the scale set by the baseline separation of the antennas (following Equation \ref{interRES:eqn}). So the greater the number of antennas in an array, the greater amount of information can be recovered about the target. Plus the greater the number of unique baseline lengths the larger the number of spatial scales your interferometer can sample in the target object.

For interferometers built with a small number of antennas, e.g. the Australia Telescope Compact Array (ATCA) \cite{ATCAWebsite} with 6 antennas or e-MERLIN \cite{eMERLINWebsite} with 7 antennas, the number of $u, v$ points attainable in an instantaneous observation is very small.  To overcome this a technique known as \textit{Earth Rotation Synthesis} can be employed. As the Earth rotates the antennas in an array will effectively rotate around the vector \textbf{s} observing new $l, m$ points ergo a new $V(u,v)$ measurement. By making multiple short observations of the same object at different times throughout a days observation, the amount of unique $u, v$ points can be increased thus gaining more information on the target object. 

\begin{figure}[!h]
\begin{center}
\includegraphics[scale=0.7]{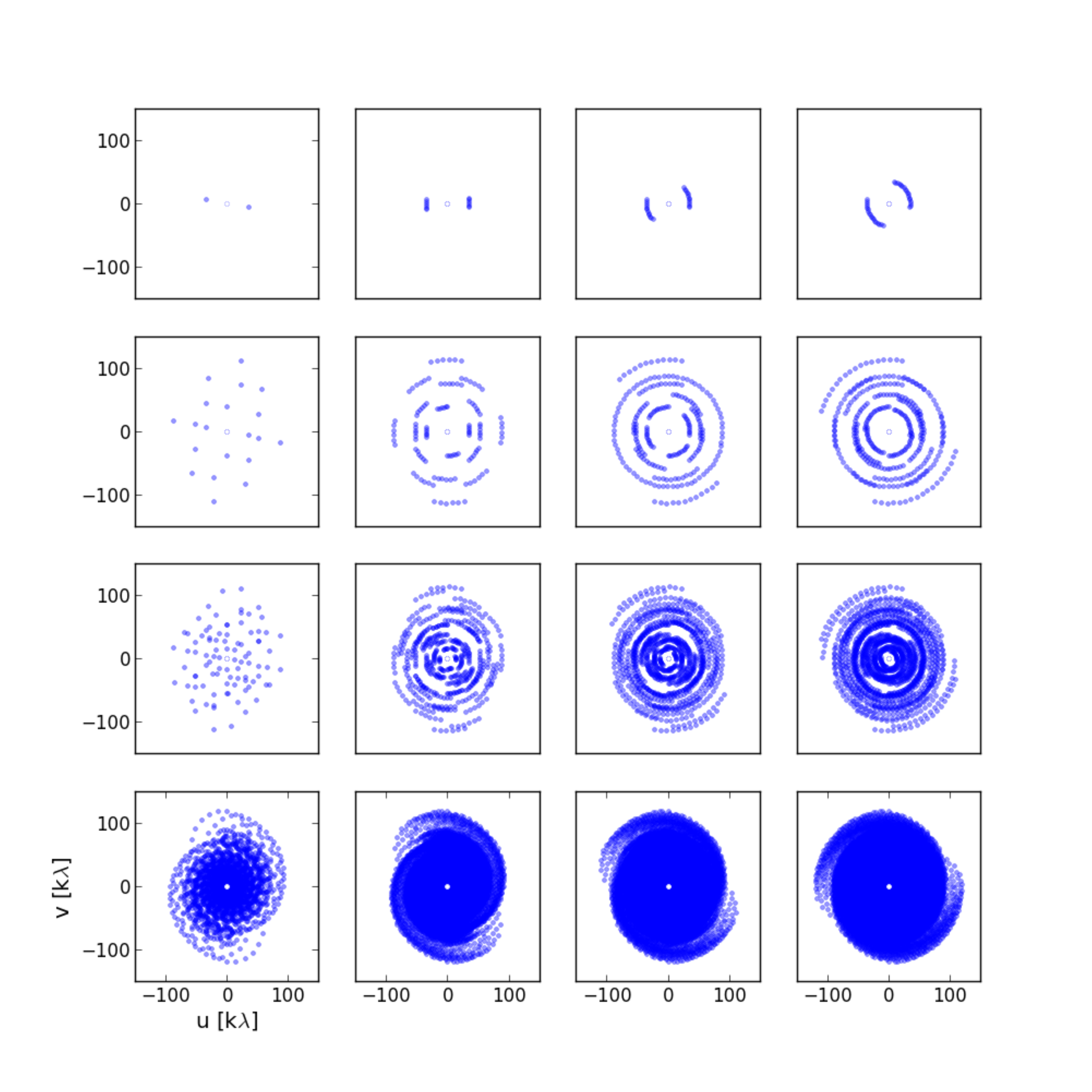}
\caption{\textit{u,v}-coverage of an interferometer set out in a logarithmic spiral pattern comprised of 2, 5, 10 and 5 antennas (top to bottom) and observing for 10 seconds, 2, 4, and 6 hours (left to right).}
\label{antstime:fig}
\end{center}
\end{figure}

To illustrate this each panel in Figure \ref{antstime:fig} shows the affect of increasing the number of antennas in an array  (2, 5 , 10, and 50 antennas, top to bottom) and in increasing the amount of time on source (instantenous image, 2, 4 and 6 hours, left to right). We can see from this that short observations with a large number of antennas over a large number of different baselines will achieve a greater total number of unique $u, v$ points than several hours observation time with a small number of antennas. This is the reason modern interferometers (such as ALMA, MeerKat, ASKAP and ultimately the SKA) all make use of large numbers of antennas. 

\begin{center}
***
\end{center}
\textbf{Note:} In this section, for mathematical simplicity we have avoided several steps used in the conversion of antenna signals to a sky brightness map. For an in depth and rigorous description of interferometry we again refer the reader to \cite{SynthesisImagingII} and \cite{IntroRadio} which were used extensively in the writing of this section. We also refer the reader to two key textbooks in the subject of interferometry and radio astronomy which are \cite{IntandSyn} and \cite{Tools}, both of which present a detailed discussion of interferometry.
\section{Simulator approach}

As the primary goal of the Pynterferometer is to demonstrate the affect on observation by changing the position and number of antennas within an array, a large portion of the mathematical and real world intricacies of an actual interferometer are avoided. The Pynterferometer takes two inputs; the user defined array configuration and an input image (in .png format) and will give one or two outputs depending on the version used. The `Basic' version outputs the simulated output image, whereas  the `Advanced' outputs both the simulated output image and a plot of the $uv$-coverage created by the user defined array.

\begin{figure}[!h]
\begin{center}
\includegraphics[scale=0.75]{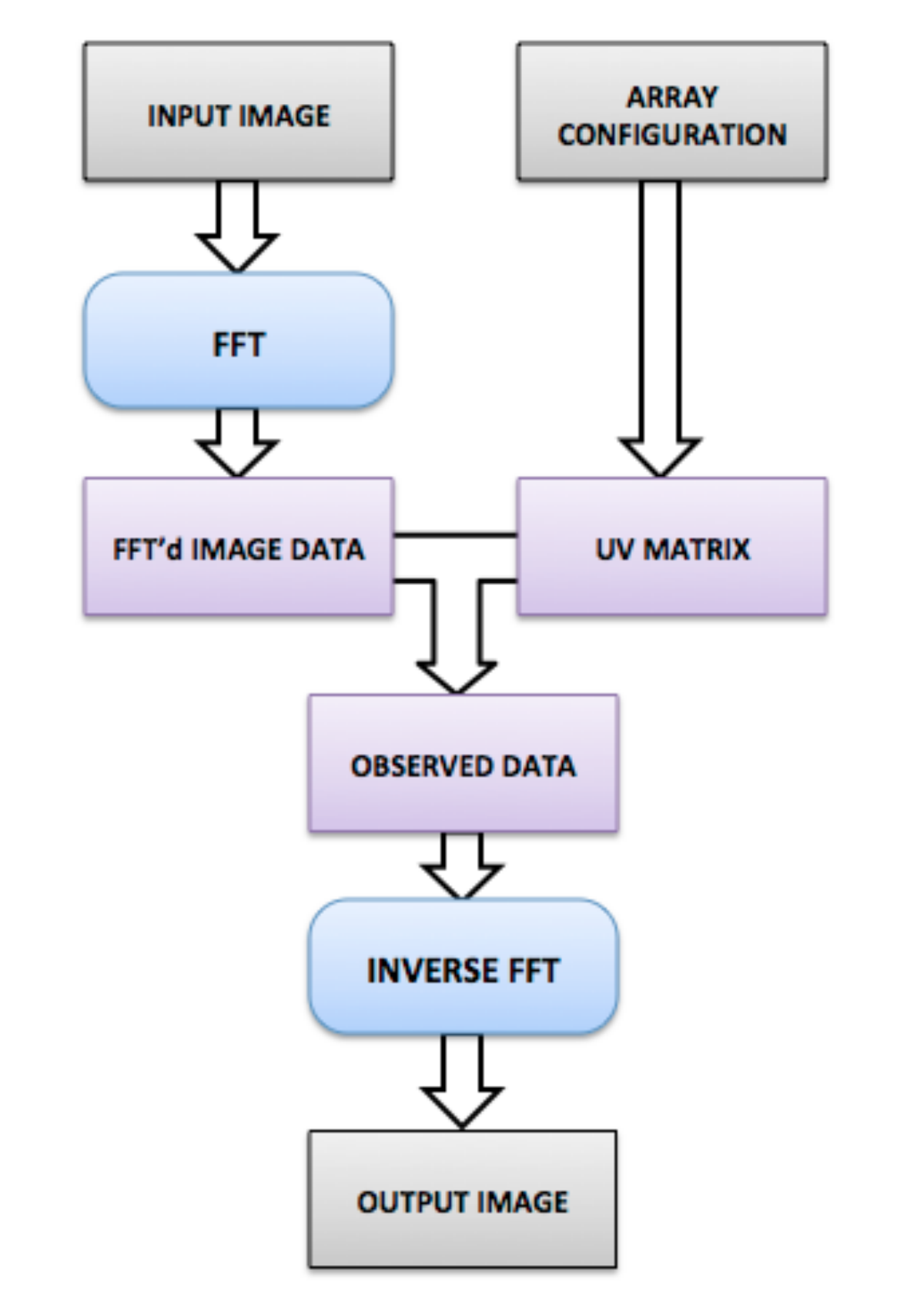}
\caption{Schematic of the dataflow and operations conducted throughout the simulation.}
\label{FFTflowchart:fig}
\end{center}
\end{figure}

Given a telescope array defined by the user, the Pynterferometer calculates the $u$ and $v$ points this array would observe and outputs these as a unit matrix. Next, the input image is Fourier transformed using a Fast Fourier Transform algorithm and the Fourier components matrix multiplied by the observed $uv$ matrix. This sets all the points in the transformed input image without a corresponding $uv$ point to zero but retains all information where there is a point. This final matrix is then inverse Fourier transformed back into the spatial domain and converted in to an image format resulting in the output image. Figure \ref{FFTflowchart:fig} shows a schematic of this process.


We make a few other assumptions with our approach that a radio astronomer would not be able to make. These are that no instrumental or atmospheric noise is included and no deconvolution of the image (as is routinely done in practice such as CLEANing \cite{DeconvolutionCornwell99}) are undertaken.

\section{The Pynterferometer}
\subsection{The interface}

\begin{figure}[!tbh]
\begin{center}
\includegraphics[scale=0.4]{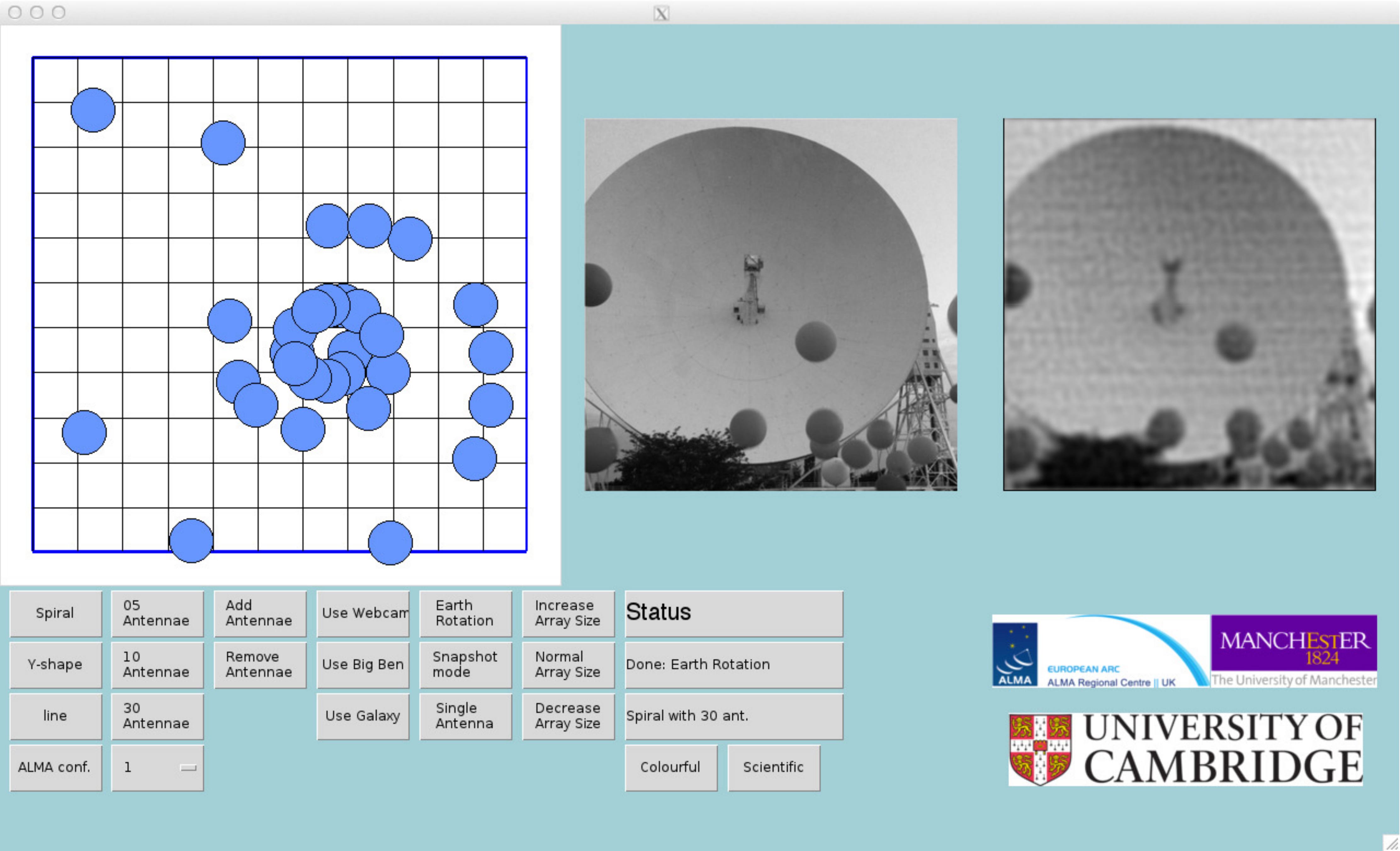}
\caption{A screen grab of the Pynterferometer running on the Mac OSX operating system. The images and buttons are described in the text.}
\label{ScreenGrab:fig}
\end{center}
\end{figure}

\begin{figure}[!tbh]
\begin{center}
\includegraphics[scale=0.4]{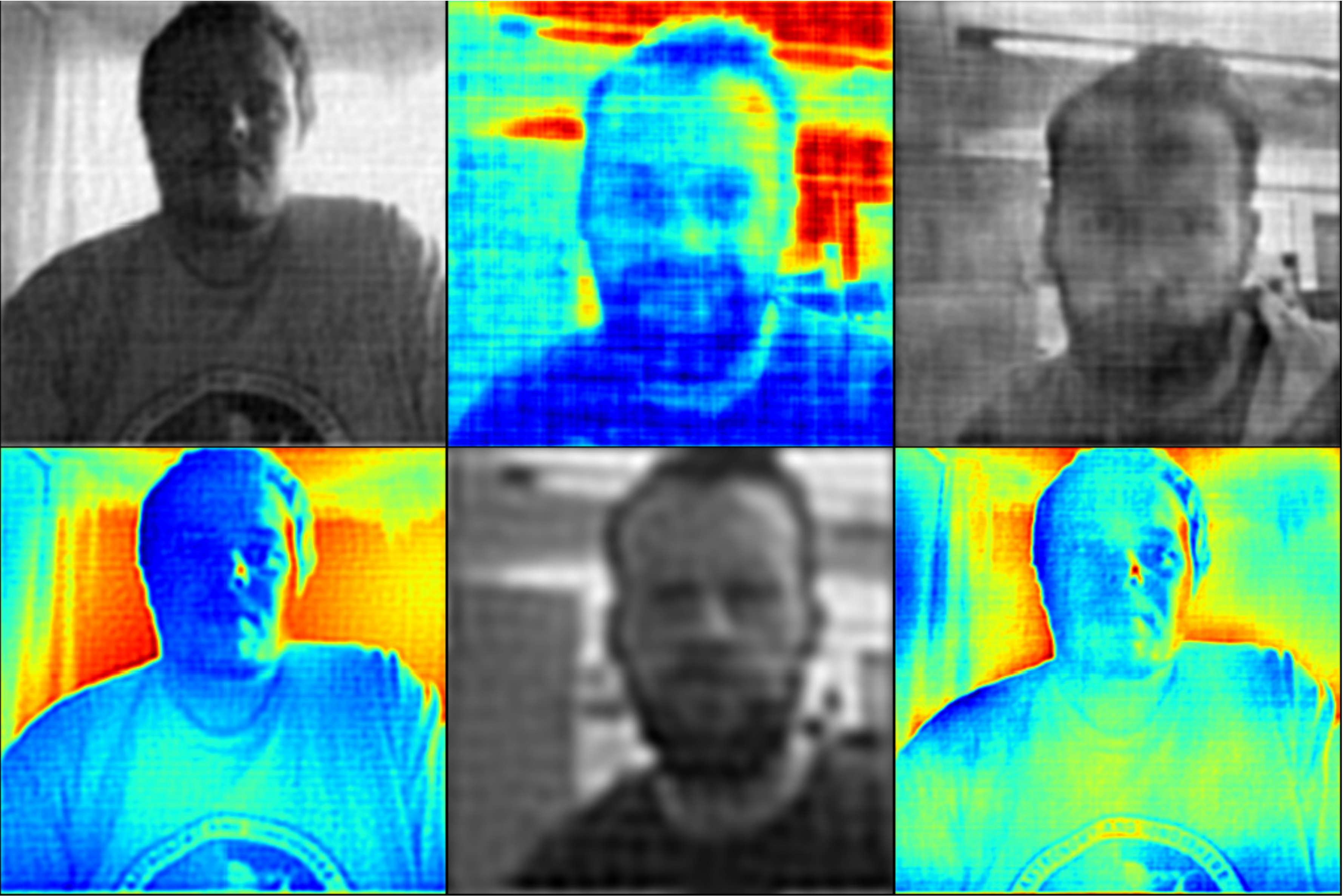}
\caption{Example Pynterferometer output images using the webcam input image mode and ``Scientific'' and ``Colourful'' colour scale settings.  Each of these six images was made with a different antenna configuration, the effect of which can be seen image to image.}
\label{mosaic:img}
\end{center}
\end{figure}

Figure \ref{ScreenGrab:fig} shows a screen grab of the basic version of the Pynterferometer. The gridded area (top left) with blue circles is the ``Array Configuration" area, users can drag the blue circles which represent antennae in the the array to any position within the grid. The array configuration in this area will then be used to calculate the output image. The input image is to the right of the ``Array Configuration" area and is the image which is used as the ``sky" image for transform. The output image (top right) is the resulting image of the transform using the input image and the defined array configuration. Example output images can be seen in Figure \ref{mosaic:img}, each image  was made with a different antenna configuration.

The lower portion of the Pynterferometer consists of a number of buttons, these (in columns from left to right) are 1) preset antenna configurations, 2) preset antenna numbers, 3) antenna addition and removal buttons, 4) image selection, either to use a webcam or one of the two image presets, 5) observing mode selection and 6) antenna area selection. Finally on the right are 3 status panels which describe what the Pynterferometer is doing (i.e. calculating $uv$ coverage) and the current number of antenna and the original configuration of the array. Two final buttons exist below the status boxes ``Scientific'' and ``Colourful'' which change the colour scale used in the output image. The bottom right of the Pynterferometer then includes the logos of the home institutions of the authors.


\subsection{System Requirements and Availability}
The Pynterferometer works on Windows 7 and Vista, Mac OSX and Linux operating systems provided the appropriate Python modules and libraries are provided. The prerequisites are packaged up with the Pynterferometer and can be downloaded from \texttt{http://www.jb.man.ac.uk/pynterferometer/}.

\section{Example demonstrations}

The Pynterferometer can be used to quickly demonstrate the effects of antenna configuration/numbers/separation on the output image. Whilst we provide these two examples,  there are many ways in which the public or a demonstrator can interact with this software to learn/teach about the imaging capability of interferometric telescopes.

\subsection{Example 1: Number of antennas}
Using the Add/Remove antenna buttons you can show how increasing the number of antennas improves the output image.
\begin{enumerate}
\item First click `Spiral' and `Snapshot'
\item Click `Single Dish'. The output image will become a blob (a 2D Gaussian convolved with the input image in
reality).
\item Click `Add antenna', the output image will then become a series of stripes. Next, increase the number of antennas using the `Add antenna' button or the presets `05/10/30 Antennas',  the image will slowly improve in quality.
\end{enumerate}

\subsection{Example 2: Array Separation}
Using the `Decrease Array Size' , `Normal Array Size' and `Increase Array Size' buttons you can show how a larger antenna separation will observe more of the small scale structure, whilst a closely packed configuration will see only the extended structure.
\begin{enumerate}

\item First click `Use Galaxy' and `Earth Rotation'.
\item Click `Decrease Array Size', the grid lines in the array panel will increase in separation, these grid lines \textbf{always} show boxes of 500m $\times$ 500m across so you are looking at a small area of land\footnote{Please note the telescope circles are oversized for ease of use.} . The output image will show the galaxy fairly well but most of what is seen is the extended/large scale structure (e.g. the overall shape of the galaxy).
\item Click `Normal Array Size' the grid lines will become more closely packed so you are looking at more land (i.e. the antennas are now further apart from one another with max separation = 7km). An increase in the amount of smaller scale structure seen in the output image will now be seen.
\item Finally, click `Increase Array Size' again the grid lines will become more closely packed so you are looking at yet more land (i.e. the antennas are now even further apart from one another; max separation = 14km). Here the output image will be devoid of most extended structure and dominated by fine small scale structure.
\end{enumerate}

\section{Reception of the tool by the public}

This tool was originally created for use on the ALMA display at the Royal Society Summer Science Exhibition 2012, which ran from the 3$^{rd}$ to the 6$^{th}$ of July 2012. The event saw 11,090 visitors during the 6 days it was open with a large proportion passing and speaking to the staff on the ALMA stand. Not all stand visitors used the tool but we estimate over a thousand uses over the whole event. 

During the exhibition there was a twitpic\footnote{twitpic.com} feature, which posted the users image to the URL \texttt{http://twitpic.com/photos/alma\_sse/} was included. Over the exhibition week the `twitpic' feature, saw $\sim$200 uses. Please note this feature is not available in the main Pynterferometer release. 

Some members of the public were happy to use the tool unsupervised and asked questions when they felt it necessary, whilst the majority were lead through the tool by one of the stand staff. The staff found this as a very useful tool for visually demonstrating the process of radio interferometry. It was used quite heavily when trying to overcome the difficulties of exampling why the Earth rotation helps.

\section{Summary}
We have presented a graphical interface designed to demonstrated the techniques of radio interferometry in a manner accessible to the general public. The Pynterferometer demonstrates how the number of antenna, their position relative to one another and the rotation of the Earth allow astronomers to create highly detailed images at long wavelengths. The aim of the Pynterferometer is to introduce the observational technique of interferometry which will become increasingly prevalent in professional astronomy in the next few years as e.g. ALMA, ASKAP, MeerKAT and the SKA all become fully functioning instruments.

{\small{\section*{Acknowledgements} The authors would like to thank Dr John Richer who lead the ALMA team at the Royal Society Summer Science Exhibition 2012 and thought this would be a good idea! The whole team who manned the stand at the Exhibition, the staff of the UK ARC Node and PDRAs and PhD students from both Manchester and Cambridge who tested this software. The Royal Society for selecting the ALMA team to display at the Exhibition. The STFC as well as the University of Cambridge, The Kavli Institute for Cosmology and the UK ARC Node for funding the ALMA Exhibition.}}\\

\bibliographystyle{unsrt}
\bibliography{AvisonGeorgePySim_Bib}

\end{document}